\documentclass[a4paper, authoryear, 12pt]{article}
\usepackage[utf8]{inputenc}

\usepackage[english]{babel}
\usepackage{pifont}
\usepackage{color}
\usepackage{babel}
\usepackage{fontenc}
\usepackage{graphicx}
\usepackage{natbib}
\usepackage[colorlinks=true,citecolor=blue]{hyperref}
\usepackage[usenames,dvipsnames]{pstricks}
\usepackage{amsmath,amsthm, amsfonts, amssymb, mathrsfs,hyperref}
\usepackage{epsfig}
\usepackage{amsthm}
\usepackage{lineno}
\usepackage{multicol}
\newtheorem{theorem}{Theorem}
\usepackage{authblk}
\usepackage{float}
\usepackage{caption}
\usepackage{subcaption}
\usepackage{cleveref}
\usepackage[toc,page]{appendix}
\usepackage{setspace}
\usepackage{appendix}
\numberwithin{equation}{section}

\providecommand{\keywords}[1]
{
  \small	
  \textbf{\textit{Keywords---}} #1
}

\title{ Interconnection between density-regulation and stability in competitive ecological network  }

\author{Amit Samadder, Arnab Chattopadhyay, Anurag Sau, Sabyasachi Bhattacharya \footnote{Agricultural and Ecological Research Unit, Indian Statistical Institute, 203, B.T Road, Kolkata- 700108, India. E-mail: math.amitsamadder18@gmail.com, arnabchatterjee891@gmail.com, anuragsau@gmail.com, sabyasachi@isical.ac.in}
\footnote{Corresponding Author, Contact No. - (+91)-9433897120, Fax:(91)(33) 2577-3049}}

\begin{document}
\date{}
\maketitle
\begin{abstract}
In natural ecosystems, species can be characterized by the nonlinear density-dependent self-regulation of their growth profile. Species of many taxa show a substantial density-dependent reduction for low population size. Nevertheless, many show the opposite trend; density regulation is minimal for small populations and increases significantly when the population size is near the carrying capacity. The theta-logistic growth equation can portray the intraspecific density regulation in the growth profile, theta being the density regulation parameter. In this study, we examine the role of these different growth profiles on the stability of a competitive ecological community with the help of a mathematical model of competitive species interactions. This manuscript deals with the random matrix theory to understand the stability of the classical theta-logistic models of competitive interactions. Our results suggest that having more species with strong density dependence, which self-regulate at low densities, leads to more stable communities. With this, stability also depends on the complexity of the ecological network. Species network connectance (link density) shows a consistent trend of increasing stability, whereas community size (species richness) shows a context-dependent effect. We also interpret our results from the aspect of two different life history strategies: r and K-selection. Our results show that the stability of a competitive network increases with the fraction of r-selected species in the community. Our result is robust, irrespective of different network architectures.
\end{abstract}


\section{Introduction}

The intrinsic mechanisms behind the stabilization of an ecological community have been a common query of ecologists since the pioneering work of May \citep{may1972will} around fifty years ago. He revealed that complexity should be a regulating factor for the stability of an ecological system. Henceforth, the complexity and stability of ecological communities is a long-standing unsettled issue and still
inconclusive. Various debates and cross-debates exist \citep{mccann2000diversity, elton2000ecology, nunney1980stability, gardner1970connectance, tilman1994biodiversity,landi2018complexity,goodman1975theory, mccann2000diversity, pimm1984complexity} regarding the issue of whether complexity is inversely proportional to stability or not. In addition to this, numerous studies elaborate on other factors, such as
adaptive foraging, multiple interactions, anti-predator defence, etc., which regulate the stability of complex communities  \citep{kondoh2003foraging,kondoh2007anti,mougi2012diversity,yan2014specific,kawatsu2018density}. Existing studies related to multi-species dynamics assume the Lotka-Volterra competitive model to demonstrate the complex, competitive ecological network \citep{kokkoris1999patterns,rozdilsky2004effects,rozdilsky2001complexity,fowler2009increasing}. The simple density-dependent logistic law regulates the individual species dynamics in the absence of interspecific interactions under this Lotka-Volterra model setup.
\newline

In the context of our study, naturally, two questions arise. (i) What should be the basis for assuming the degree of density-dependence to explain individual growth dynamics? (ii) Is logistic growth the best candidate to explain the density-dependent phenomenon in reality? Searching for the answer to these questions will motivate us to form our conceptual model. Let us discuss this briefly in the following paragraph.   
\newline

Density dependence, population regulation, and variability in population size are three yardsticks and debated concepts of population ecology. The operation of some regulatory processes is associated with low variability. This regulatory process involves one or more density-dependent components \citep{hanski1990density}. The degree of density dependence variability is negatively correlated with the average variability level, suggesting that the more regulated populations are generally less variable. Moreover, \cite{hanski1990density} pointed out that density-dependent regulation and variability are more prominent for species with complex dynamics in nature. Most growth law-associated density-dependent factors are biotic, including predation, inter and intraspecific competition, accumulation of waste, and diseases such as those caused by parasites. Other density-dependent factors are abiotic. So, it is evident that the growth rate and density-dependent relationship is usually complex due to several biotic and abiotic factors. So it is hardly reasonable to assume a linear decline of per-capita growth rate (pgr) via an instantaneous response to current density, which is the basic assumption of the logistic equation. The density-pgr relationship is preferable to be skewed over the standard logistic setup \citep{schoener1973population,mallet2012struggle,roughgarden1997production}. Logistic growth can rarely be observed in real-life phenomena, and hence unrealistic. So, theta-logistic with a nonlinear density-pgr relationship is a reasonable choice in this case.  
\newline

Theta-logistic model is very popular even today and hence extensively used to explain the dynamics of various ecological \citep{eberhardt2008analyzing,paul2022estimation} and epidemiological \citep{wang2012richards} processes. In some cases, parameter estimates are unreliable and unrealistic \citep{ross2009note, clark2010theta}. These implausible estimates can be avoided if we use the profile likelihood estimates \citep{polansky2009likelihood,paul2022estimation} instead of the usual grid-search and least square techniques. 
\newline

Let us return to the mainstream again, where we are mainly interested in understanding the stability of the complex competitive network where the species randomly interact with each other. The species growth is density-dependent, and the density-pgr relationship is linear in nature. How the species coexistence and persistence as a notion of stability of the above system change with the change of complexity were elaborately studied by \cite{rozdilsky2001complexity}. \cite{allesina2012stability} also analyzed the stability criteria of the complex multi-species ecological network under three scenarios - prey-predator, mutualistic, and competitive models. Several vital issues emerge if the species experiences non-linear density-dependent growth instead of regular logistic mechanism. (i) Under the theta-logistic model setup, how do the local community stability and resilience pattern change with changing network complexity? We interpret complexity by network connectance and
species richness. (ii) It might be an exciting research finding for ecologists to understand how strong density regulations at low, high population density influence the stability of multi-species competitive networks with random interactions. Finally, the classification of r and K-selected species based on the density-regulation parameter (defined as $\theta$ in the next section) may be another prime aspect that must be explored. In this connection, \cite{clark2010theta} suggested that concave r-N curves are typical of so-called r-selected organisms where density-dependence acts strongly at lower densities. In contrast, convex curves arise from K-selected species where density dependence acts to reduce growth only at higher densities. The parameter theta reflects the evolved life history pattern in terms of demographic rates that
determine how abruptly growth slows as abundance interacts with resource availability and the type of competition \citep{clark2010theta}. 
\newline

In this paper, our goal is to investigate the effect of strong intraspecific density-dependence at low(high) population size on a competitive ecological community with random interactions between species. To do this, we generalize the Lotka-Volterra model by considering non-linearity in species’ relative growth rate profile using the theta-logistic model. We hypothesize that non-linear density dependence may significantly impact maintaining stability and resilience in a competitive community. Furthermore, we investigate the role of network complexity on stability. We derived a theoretical condition for stability in the case of random network structure with some relaxed assumptions and validated it by performing extensive numerical simulations. Other network structures, viz., scale-free and small-world networks, are also considered for the simulation\citep{barabasi1999emergence,newman2018networks}. At the same time, just like several previous researchers \citep{clark2010theta, paul2022estimation, pianka1970r, bhowmick2015cooperation}, we attempt to establish the relation between r-K status and density regulation parameter, although it involves several debates.  
 \newline
 
 The paper is organized as follows. Section \ref{theta-logistic} contains a brief discussion about theta-logistic model. Section \ref{Materials and methods} contains the mathematical and biological concepts behind the formulation of our model. Then, we also analyze the local stability through the random matrix theory in this section. In section \ref{Results}, we simulate our result through numerical technique. In section \ref{Discussion}, we discuss some important results and ecological aspects of our result, and finally, we end with a conclusion in section \ref{Conclusions}.
 
\section{The theta-logistic model: Origin and application, limitation, and exemption from it}
\label{theta-logistic}

 {\it \bf Origin and application:}  A different form of the theta-logistic model was first introduced by \cite{richards1959flexible} when he doubted the theoretical validity and usefulness of the Von-Bartalanffy model. \cite{gilpin1973global}  rediscovered the theta-logistic model by allowing more freedom in the form of density dependence. It is to be noted that both logistic and theta-logistic are density-dependent. However, when an additional parameter is incorporated into a simple logistic model, the density regulation or population regulation concept can be realistically captured for species growth problems. \cite{gilpin1973global} proposed the following form of density-dependent theta-logistic model: 

\begin{equation}
   \frac{dX}{dt} = rX \left(1-\left(\frac{X}{K}\right)^{\theta}\right),   
\end{equation}

 where $X$ is the population density, $r$ is the intrinsic growth rate, and $\theta$ is the non-linear effect of con-specific density on the growth rate. The parameter $\theta$ controls the shape of the growth curve. This model reduces to the standard logistic growth curve when $\theta=1$. When $\theta <1$, density-dependence is strong at low population density. By contrast, density-dependent is weak when $\theta>1$
until the population is close to the carrying capacity \cite{clark2010theta}. In natural systems, the parameters $\theta$ and $r$ are not independent but rather exhibit a strong negative correlation\citep{clark2010theta,saether2008forms,saether2002demographic,fowler1988population}.
\newline

The growth mechanisms of several species have been well explained through this theta-logistic model. For example {\it acorn woodpecker} and several other populations of birds experienced theta-logistic pattern in their growth process \citep{stacey1992environmental, saether2000estimating, saether2002pattern, saether2008forms, ross2009note}. Evidence of theta-logistic growth profiles is also prominent for other species like sea otters({\it Enhydra lutris L.}) off the coast of Washington State, U.S.A. \citep{gerber2004density}, and elephant ({\it Loxodonta africana Blumenbach}) in Hwange National Park, Zimbabwe \citep{chamaille2008resource}. 
\newline

\cite{wang2012richards} revisited the \cite{richards1959flexible} (later known as the theta-logistic model) model and identified its application in infection dynamics. This model is validated by four sets of real data and provides reliable estimates in predicting disease outbreaks and detecting the turning point and multiple weaves or phrases. \cite{paul2021covid} estimated that the penalty of epidemiological models outcast the simple population dynamic solution for estimating the outbreak COVID-19 pandemic. The author used the theta-logistic model rigorously and robustly to predict the different phrases and decrease-eradication time frame. More recently \cite{paul2022estimation} theoretically estimated the environmental intensity a species could tolerate and validated through a large scale of GPDD data. The author used the stochastic theta-logistic model and estimated the probability of extinction for species of four taxonomic groups (viz. birds, mammals, bony fish, and insects). 
\newline

{\it \bf Limitations:} The \cite{sibly2005regulation} undertook an ambitious analysis of this relationship by examining growth rates of 674 species of four taxonomic groups and concluded that most of the density-pgr relationships are concave in reality. Sibly's study elicited a series of debates emerging from different perspectives. For example, \cite{getz2006comment} raised some adverse technical comments, specifically in model selection, interpreting empirical results, and dismissing trophic interactions and environmental stochasticity. Ross's \citep{ross2006comment} concern revolved around the physical interpretation of the model parameters. Finally, \cite{peacock2006comment} pointed out the distortion of Sibly's conclusion for harvest data. Moreover, after a span of $5$ years, \cite{clark2010theta} raised the concern on reliability of the theta-logistic model for modeling census data and finally refuted the claim of \cite{sibly2005regulation} that concave growth responses are shown to dominate in nature. He suggested that the abundance data is generally associated with fluctuating natural populations. The profile of steady growth movement from low abundance is hardly visible in such data. Thus, we can only measure part of the entire growth trajectory. The actual dynamics of a natural population over each time step will often differ profoundly from the tendency suggested by the
model because these phenomenological models describe only a population's tendency to return to the carrying capacity. \cite{wang2012richards} claim that fitting theta-logistic model to some data evolves some extreme and biologically implausible estimated parameter values due to over-fitting problems. This issue can be handled by reducing the parameter using suitable constraints. There is a chance for estimating negative parameter values for the theta-logistic model if we use either grid-search or non-linear least square techniques. 
\newline

{\it \bf Exemption:} To overcome this limitation, the profile likelihood method is the best choice, especially for population census data \citep{polansky2009likelihood, paul2022estimation}. Moreover, a slight fluctuation around carrying capacity may substantially increase the percentage of the concave density-pgr relationship. However, this can be avoided if the measurement error is included or stochasticity is inbuilt into the system's appearance (random interaction of competitive species) by definition.  \cite{clark2010theta} harshly used the word "unreliable" for the theta-logistic model in dealing with census data. We are a bit conservative in using the word "unreliable." We claim that the theta-logistic model is the simplest non-linear growth model, and the contribution of the model is undeniable. Nevertheless, at the same time, while using the theta-logistic model for census data, caution should be exercised in selecting the estimation procedures.

\section{Material and methods}
\label{Materials and methods}

\subsection{Mathematical Model}
\label{Mathematical Model}

Several competitive species govern the complex ecological community. Intraspecific and interspecific interactions in a competitive community can play a crucial role in species coexistence. The chance of species coexistence increases if the intraspecific effect should be greater than the inter-specific effect in two species dynamics  \citep{chesson2013species,adler2018competition}. The intraspecific competition is governed by the linear density dependence. Let us assume the growth profile for an $N$-species community whose dynamics are driven by competitive species interactions. In the complete competitive system, we delineate two components. The first component is dedicated to the dynamics of individual intraspecific growth, while the second component focuses on interspecific dynamics. The intraspecific competition is governed by density regulation, which is represented by the theta-logistic model in the absence of neighboring competitors, as concluded by \cite{sibly2005regulation} (which is elaborately mentioned in Section \ref{theta-logistic}). For interactions involving non-identical species, we presume linear density-dependent interspecific competition. This assumption of linear interactions among competitors aligns with the perspectives of various authors, such as \cite{fowler2009increasing,rozdilsky2001complexity}. Thus, the complete competitive system based on intra and interspecific equations is defined using the following ordinary differential equations

\begin{equation} \label{final_equation}
 \frac{dX_{i}}{dt} =G_{i}= r_{i} X_{i} \left(1-\left(\frac{X_{i}}{K_{i}}\right)^{\theta_{i}}\right)+\sum_{j=1,j\neq i}^{N} a_{ij} X_{i} X_{j},
 \end{equation}
 
 This model can be reparameterized as, 
 
 \begin{equation} \label{final_equation_reparameterized} 
 \frac{dX_{i}}{dt} =G_{i}=  X_{i} \left(r_{i}-\left(a_{ii}X_{i}\right)^{\theta_{i}}\right)+\sum_{j=1,j\neq i}^{N} a_{ij} X_{i} X_{j},
 \end{equation}
 
where $X_i$ is the abundance, $r_{i}>0$ is the intrinsic  growth rate, and  $K_{i}$ is the carrying capacity of the $i-th$ species   $(i=1,...,N)$. $\theta_{i}$ is the parameter controlling the shape of the density-dependent growth curve of the $i$-th species; $\theta_{i}<1$ ($\theta_{i}>1$) for strong (week) density-dependent self-regulation at low population density. If $\theta_{i}=1$, the above model is equivalent to the logistic growth equation. Here, $a_{ii}=\frac{r_{i}^{1/\theta_{i}}}{K_{i}}$ is the self-interaction coefficient for identical species, while $a_{ij}(<0)$ is the cross-interaction coefficient between any two non-identical species within the community.
\newline

For a random model of community structure, a link between any pair of species $i,j (i, j = 1, ..., N )$ occurs with probability $C$ (connectance) and the maximum number of links $L_{max}$ is given by $\frac{N(N-1)}{2}$ \citep{newman2018networks}. In general, the food web is composed of a hierarchy structure, while mutualism is formed in a bipartite manner. However, there are very few restrictions on a competitive structure. In the natural ecosystem, any structure may be possible for a competitive network. In addition to the random network, we investigate two other types of network structures, viz., scale-free and small-world networks. In a scale-free model, most species have few competitive partners, but some species interact with large numbers of species. The degree distribution of the network nodes follows the power-law $P(k)\propto k^{-\gamma}$, where the probability of finding a node of degree $k$ is $P(k)$, $\gamma$ is the power-law exponent \citep{barabasi1999emergence, newman2018networks,servan2018coexistence}. A small-world community structure can be viewed as the interpolation of regular and random structures. Regular networks consist of interactions among neighboring nodes with an equal degree of each node, which is then rewired with probability $q$ between randomly selected nodes. If $q = 0$, it represents a completely regular network, while $q = 1$ gives a random network. The small-world network is somewhere for $0 < q < 1$, $q$ being sufficiently small, maintaining a high clustering coefficient and low average path length \citep{watts1998collective}. \newline

\subsection{Analytical solution for local stability}
\label{Analytical solution for local stability}

Analytic behaviors of a system can easily be assessed by linearizing the system near a feasible equilibrium point. The Jacobian matrix corresponding to this simplified system is also known as the community matrix ($M$). The diagonal elements of the community matrix $M$ can be obtained as partial derivatives of the growth equation of species $i$ with respect to the abundance of itself at equilibrium. Hence, for the equation \ref{final_equation_reparameterized}, the community matrix at the equilibrium point is given by,

\begin{equation}
\label{diagonal elements_community matrix}
M_{ii}={\frac{\partial G_{i}}{\partial X_{i}}|_{{X_{i}}^{*}}}=-{\theta_{i}}{(a_{ii}{X_{i}}^{*})}^{\theta_{i}},
\end{equation}

where  $X_{i}^{*}$ denotes the equilibrium density of the species $i$ ($X_{i}^*>0$). The off-diagonal element $M_{ij}$ can be obtained as the partial derivative of the growth equation \ref{final_equation_reparameterized} of species $i$ with respect to the abundance of species  $j$ at equilibrium, i.e.,

\begin{equation}
M_{ij}={\frac{\partial G_{i}}{\partial X_{j}}|_{{X_{i}}^{*}}}=a_{ij}X_{i}^{*}.
\end{equation}

This system is locally stable if all eigenvalues of the community matrix $M$ have negative real parts. The analytical condition of local stability for a small system can be easily obtained by finding the general expression of a fixed point for a given parameter space. When the system is too large, we cannot analyze the system through this method. Moreover, it is hard for a given parameter set to determine the numerical values of the fixed points in the presence of the nonlinear density-dependent parameter in the per-capita growth rate. Hence, we consider the random matrix theory (RMT) approach, introduced by May in 1972 \citep{may1972will}. In this seminal article, the author considered a random community matrix $M$ of size $N\times N$, which is the Jacobian at the equilibrium point of a random ecosystem containing $N$ species. The element $M_{ij}$ of the matrix $M$ indicates the effect of species $j$ on $i$ at the equilibrium point. The author assumes that the diagonal elements of the community matrix $M$ are all $-1$, and the off-diagonal elements are sampled from a normal distribution $N(0,\sigma^2)$ with probability $C$ and are equal to $0$ with probability $1-C$. For a large number of species, the author justified that the system is likely to be unstable whenever the "complexity" $\sigma \sqrt{NC}>1$. On the contrary, some authors considered linkage density ($C$) and species richness ($N$) as a measure of complexity \citep{pimm1984complexity, landi2018complexity}. Although his work is based on an arbitrary ecological community, it was further extended for particular interaction types and more complex ecosystems (e.g., prey-predator, competitive, mutualistic) by \cite{allesina2012stability} with the RMT approach. They found remarkable differences between the prey-predator and competitive-mutualistic interactions; the stability increases for the prey-predator interactions and decreases for the other two cases. A substantial number of researchers used the RMT approach to analyze a random ecosystem's stability in their subsequent studies \citep{kawatsu2018density,mougi2012diversity, mougi2014stability,barabas2017self, gibbs2018effect}.  
\newline

We consider a random competitive community matrix $M$ with connectance $C$, the elements $M_{ij}$ are taken from half-negative distribution $-|\mathbf{X}|$, where $E(\mathbf{X})=0$, $Var(\mathbf{X})=\sigma^{2}$. Then, $E(M_{ij})_{i\neq j}=-CE(|\mathbf{X}|)$, $Var(M_{ij})_{i\neq j}=C\sigma^{2}-C^{2}E^{2}(|\mathbf{X}|) $\citep{allesina2012stability}. For simplify, we set  $K_{i}=1$, $X_{i}^{*}=X^{*}$, $\theta_{i}=\theta_{1} (< 1)$, $\theta_{2} (>1)$ and $M_{ii}=d_{1},d_{2}$ (for $\theta_{1}$ and $\theta_{2}$, respectively),  with proportions $NP$ and $N(1-P)$, respectively. Note that, for the local stability criteria, the expected spectral abscissa ($\eta(M)$) should be negative. \newline

One of the main assumptions behind the analytical derivation of the stability criterion of our proposed model is that the equilibrium population densities are all equal. The analytical derivations of the random matrix approach are recently available in \cite{gibbs2018effect} for unequal equilibrium sizes. However, a straightforward application of this method in the case of our model is challenging. To use the method of \cite{gibbs2018effect}, the community matrix $M$ must be of the form $M = X A$, where $A$ is an elliptic random matrix. However, in this case, $A$ can not be written as an elliptic matrix due to the modified diagonal entries of equation \ref{diagonal elements_community matrix}. Thus, we maintain the assumption of equal equilibrium sizes for the analytical results and proceed to extend the concept numerically to account for unequal equilibrium densities.\newline

The expression of $\eta(M)$ (see Appendix \ref{Appendix:A} for the derivation) is given by

\begin{equation}
\begin{split}
   \eta(M) &
   =x{\sqrt{N(C\sigma^{2}-C^{2}{E^{2}}(|\mathbf{X}|))}} +d_{2}\\
   & +\frac{{\sqrt{N}}C(1-C){E^{2}}(|\mathbf{X}|)}{{\sqrt{C\sigma^{2}-C^{2}{E^{2}}(|\mathbf{X}|)}}}\left({\frac{P}{x+\frac{\Delta}{\sqrt{N(C\sigma^{2}-C^{2}{E^{2}}(|\mathbf{X}|))}}}}+\frac{1-P}{x}\right)+CE(|\mathbf{X}|), 
\end{split}
\end{equation}

 for every $0\leq P\leq1$ and $x$ satisfy
 
 \begin{equation}
  \frac{P}{(x+\frac{\Delta}{\sqrt{N(C\sigma^{2}-C^{2}{E^{2}}(|\mathbf{X}|))}})^{2}}+\frac{1-P}{x^{2}} =1,
\end{equation}

where $\Delta=d_{2}-d_{1}>0$.
\newline

Resilience is one of the important metrics related to community stability. It measures how fast a locally stable system returns to its equilibrium after a small perturbation. It is calculated as the absolute value of the largest real part of the Jacobian matrices $M$ of a stable system at equilibrium, i.e., resilience=$|{max\{Re \lambda_{i}(M): i=1,2, . . . , N\}}|$. We also measured resilience in our numerical simulation.

\section{Numerical simulations and results} 
\label{Results}

It is known that when the density regulation is strong at low density, $\theta$ must be less than one. On the other hand, $\theta$ is greater than one for the weak density regulation at low density. It is of paramount interest to understand how community stability will be affected if we increase the proportion of strong density-regulated species in the community. We define the proportion of strongly density-regulated species by the parameter $p$. We gradually increase the parameter $p$ and observe its effect on community stability. Assigning some fixed value of $p$ indicates the community is classified by $100\times p \%$ of strong density regulated and $100\times(1-p) \%$ of weak density-regulated species. $\theta$ must be less than one for a $100\times p \%$ of species that self-regulated strongly at low density, whereas $\theta$ must be greater than one for a $100\times(1-p)  \%$ of species. In general, when the community size is $N$, we assign random $\theta$ values for $N\times p$ and $N\times (1-p)$ numbers of strong and weak density-regulated species, respectively. The random $\theta$ values are chosen from $U[0.1, 1]$ and $U[1, 4]$, respectively, for strong and weak density-regulated species in our simulation. \\

We need to know the structure and magnitude of interaction coefficients to derive the community matrix. For this, we first need to determine whether any two non-identical species have interactions or not. To do this, we need to adopt the following steps.\\

First, we fixed the value of $C$ from $0.1$ to $1$ with a step length of $0.1$. Now, we draw a random value from $U[0,1]$ for each pair of species $i$ and $j$. If the value is below the $C$, then we assure the presence of interactions (hence assigned $1$); otherwise, there are no interactions (assigned $0$) and formed the adjacency matrix. There may be a possibility that one or more row(s) of this adjacency matrix may be entirely composed of $0$. This implies that the associated species contributing to that row have no competitors. This is technically known as an isolated species. We excluded that network from our simulation to avoid the unrealistic phenomenon of population ecology. All the species must have at least one competitor in our simulation. If the interactions exist, the strength of the interaction, symbolically stated as $a_{ij}(i\neq j)$, can be determined based on the random sample $U[-1,0]$ (excluding $0$) in the spirit of \citep{chen2001transient}.\\

We adhere to the following steps, outlined in the subsequent lines, to analyze community stability while varying the proportion $p$. For a fixed structure, with particular parameter sets and a fixed $p$, we generate a positive vector of equilibrium $X^{*} =(X^{*}_{i})^{n}_{i=1}$, with each $X^{*}_{i}$ uniformly distributed in $(0,1)$. We set the carrying capacity $K_{i}=1$ for the entire simulation. The self-interaction coefficient is $a_{ii}={r_{i}}^{1/\theta_{i}}/K_{i}={r_{i}}^{1/\theta_{i}}$. Then a vector of intrinsic growth rates, $R=(r_{i})^{n}_{i=1}$, is chosen so that, $ r_{i}-(a_{ii}X^{*}_{i})^{\theta_{i}}+\sum_{j=1,j\neq i}^{N} a_{ij} X^{*}_{j}=0$ for all $i$ and $j$.\\

Now, we have all the known parameters and equilibrium abundance of all species to quantify the eigenvalues of the Jacobian matrix at the equilibrium. Using this approach, we determine whether a particular community is stable or not. We replicate this process $10000$ times to estimate the frequencies of a locally asymptotically stable system. The probability of local asymptotic stability is estimated through the relative frequency of locally asymptotically stable systems across all $10000$ sample competitive webs for a particular combination of the parameters $N$, $C$, and $p$. The mean resilience is calculated over the set of locally asymptotically stable systems among these same $10000$ sample communities. We set resilience to zero for the unstable systems, as the perturbed system will never reach its equilibrium even after a sufficiently long time.\\

We also took a different simulation approach related to solving the differential equations (equations \ref{final_equation_reparameterized}). In the first step of doing this, we need to specify the parameters of the growth curves. Additionally, we need to select the initial population densities for estimating the equilibrium values. Note that there is no suggested protocol for choosing the parameter $r$ as per the previous method. In the new simulation protocol, we have to select $r$ in such a way that either $r$ has some relationship with $\theta$ \citep{clark2010theta, saether2008forms, saether2002demographic, fowler1988population} or $r$ and $\theta$ are completely uncorrelated. We considered both scenarios. For the independent case, we chose $r$ from the uniform distribution, and for the correlated setup, we used the inverse relationship of $r$ and $\theta$. Numerical integration is performed using the Euler integration method for $10000$ time units with a step size of $0.01$ and arbitrary positive initial densities. Note that the initial richness of the networks is $100$, and we used $1000$ replications for each case. According to this simulation protocol, some species may become extinct at a steady state. For estimating spectral distribution, we exclude them from the community matrix.

  \begin{figure}[H]
    \centering
    \includegraphics[width=12cm]{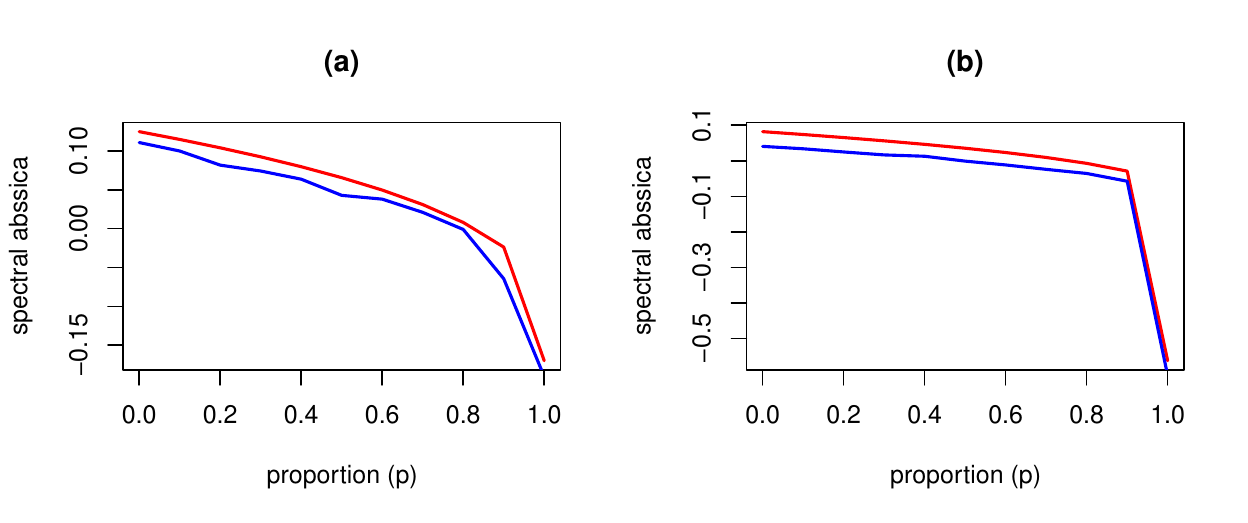}
    \caption{ Comparison between spectral abscissa (red line) and an average of the real part of leading eigenvalue (blue line) of $50$ community matrix as a measure of stability.  The proportion of species ($p$) having density regulation ($\theta_{1}$) $0.1$ is plotted in the x-axis for panel (a). It is obvious that the density regulation ($\theta_{2}$) for complementary proportion ($1-p$) is $2$. The intrinsic growth rates for the two types of species are $5$ ($r_{1}$) and $0.25$ ($r_{2}$), respectively. The community size ($N$) and connectance ($C$) are $50$ and $1$, respectively. Competitive strength ($a_{ij}$) are chosen from $U[-0.2, 0]$. The equal equilibrium size for all the species is $0.6$ ($X^{*}$). In panel (b), the proportion of species ($p$) having density regulation ($\theta_{1}$) $0.25$ is plotted on the x-axis. Density regulation ($\theta_{2}$) for complementary proportion ($1-p$) is $3$. The intrinsic growth rates for the two types of species are $3.6$ ($r_{1}$) and $0.3$ ($r_{2}$), respectively. The community size ($N$) and connectance ($C$) are $80$ and $0.9$, respectively. Competitive strength ($a_{ij}$) are chosen from $U[-0.1, 0]$. The equal equilibrium size for all the species is $0.5$ ($X^{*}$). 
    }
    \label{fig:my_label 1}
\end{figure}

\subsection{Spectral distribution and Spectral abscissa of random community matrices}
\label{Spectral distribution and Spectral abscissa of random community matrices}
  
\begin{figure}[H]
    \centering
    \includegraphics[width=12cm]{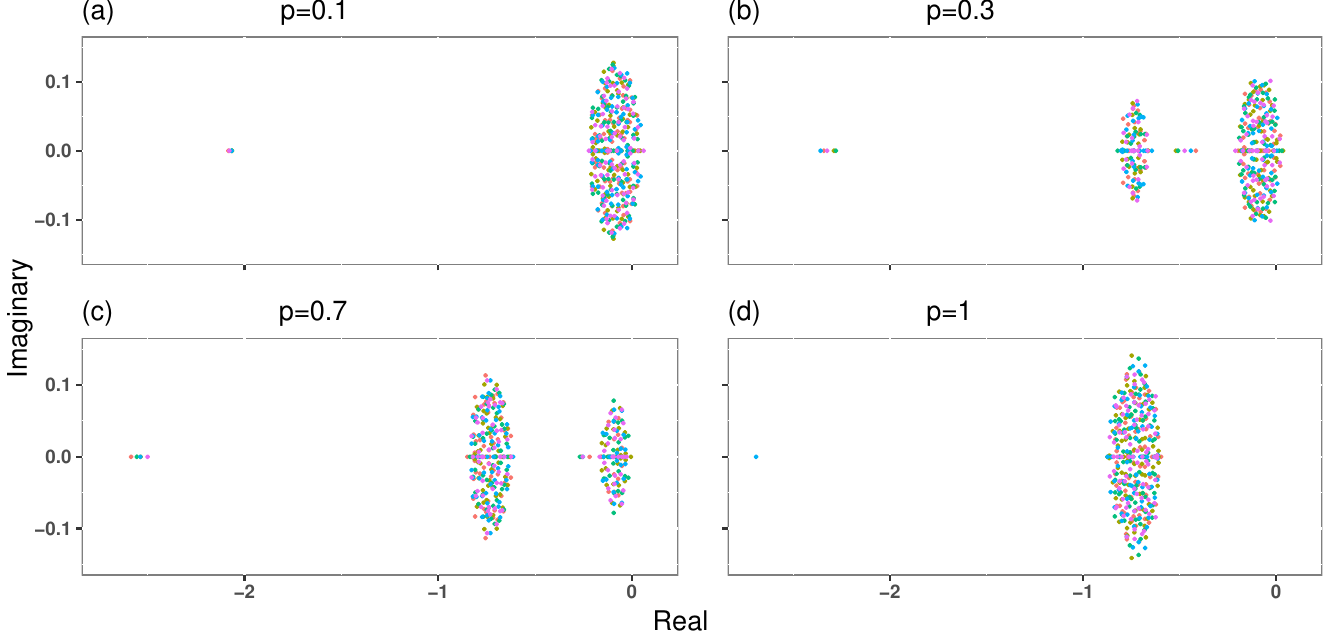}
    \caption{These panel of figures show the eigenvalue spectrum of five competitive matrices of community size $N=80$ and connectance $C=0.9$ for increasing order of proportion of strong density regulated species. We use the parameters of Fig. \ref{fig:my_label 1} for panel (b). The eigenvalue spectrum is shifting to the negative side in the complex plane with $p$ and hence increases stability. 
    }
    \label{fig:my_label2}
\end{figure}

 For an ensemble of randomly generated community matrices with fixed connectance($C$) and community size($N$), the expected maximum real part of eigenvalues is given by $\eta(M)$ (expected spectral abscissa). For stability of the underlying topology, $\eta(M)$ should be negative. Figure \ref{fig:my_label 1}(a) and \ref{fig:my_label 1}(b) both are the same type; the only difference is that they are computed for different parameter sets. These figures exhibit how the maximum real part of eigenvalues of a random competitive community matrix changes with the proportion $(p)$ of strong density regulated species (i.e., $\theta<1$). We investigate how $\eta(M)$ is regulated by connectance ($C$), community size ($N$), and the proportion $(p)$. $\eta(M)$ decreases as $p$ increases in the system for randomly generated competitive community matrices, depicted in Fig. \ref{fig:my_label 1}. The spectral abscissa flips the sign from positive to negative for a sufficient proportion, making the system stable. We verify the result with numerical simulation (Fig. \ref{fig:my_label 1}) by considering the average of maximum eigenvalues among fifty randomly generated community matrices for fixed connectance and community size. In Fig. \ref{fig:my_label2}, we illustrate how the proportion of two kinds of species ($p$) affects the eigenvalue distribution of random community structures. In the systems with strong density regulation at low or high population sizes ($p=0$ or $1$, respectively), the eigenvalue distribution appears to be elliptic. For $p$ lying between zero and one, eigenvalue distribution is subdivided into two ellipses, and the sizes of these ellipses are proportional to the proportion of weak and strong self-regulation at low population densities ($p$ and $1-p$). As the size of the ellipse close to zero increases, the length of the major axis increases, leading to eigenvalues bringing it close to zero and reducing its stability.
 
 \subsection{Numerical simulations for community stability and resilience}
 \label{Numerical simulations for community stability and resilience}
 
We aim to investigate community stability and resilience by varying the proportion $p$.  A competitive community is established based on the specified values of community size ($N$) and connectance ($C$). The process of forming a competitive community is outlined in Section \ref{Results}. We take two simulation approaches to investigate community stability and resilience, which have already been discussed in Section \ref{Results}. The first approach selects the equilibrium points from a uniform distribution. However, we did not assume any distribution for the equilibrium points in the second case. The points are generated as a solution of a differential equation with random initial conditions.
 
\subsubsection{Case \mbox{I}: Known equilibrium distribution}
\label{Known equilibrium distribution}

\begin{figure}[H]
    \hspace{-3cm}
    \centering
    \includegraphics[width=12cm]{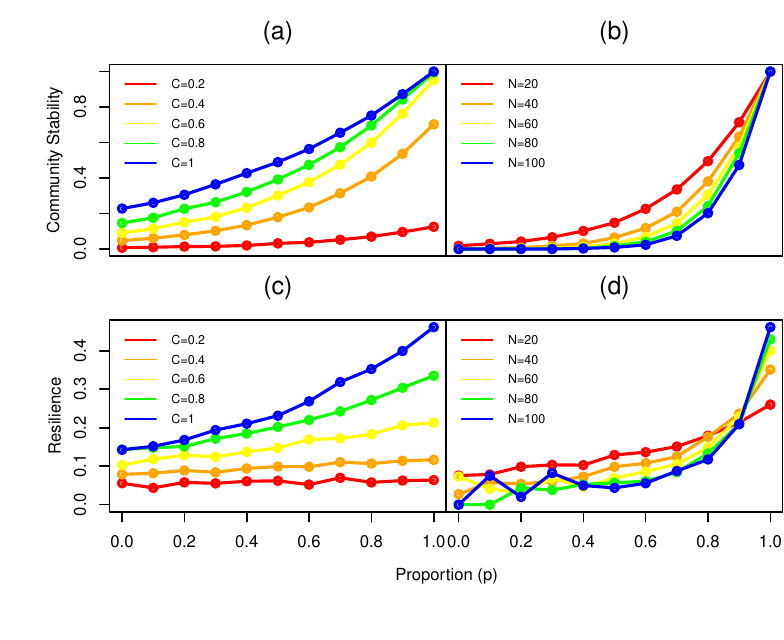}
    \caption{ Panels (a) and (b) represent the community stability profile, whereas panels (c) and (d) depict the resilience pattern. The stability and resilience pattern for fixed community size ($N=10$) and varying connectance are exhibited in panels (a) and (c). The same protocol is followed to generate panels (b) and (d) for fixed connectance ($C=0.5$) with varying species richness. Both stability and resilience have a positive association with an increase in strong density-regulated species. }
    \label{fig:my_label3}
\end{figure}

We initially fixed the value of $N$ and investigated the stability and resilience by varying $p$ while keeping the value of connectance ($C$) within a defined range. Similarly, we vary $p$ for fixed connectance ($C$) for different community sizes($N$). Figure \ref{fig:my_label3}(a) illustrates that connectance has a positive effect on stability, as stability increases with connectance for a system characterized by a fixed proportion of strong density regulation ($p$) and community size ($N$). However, the effect of community size is the opposite of connectance (Fig. \ref{fig:my_label3}(b)). Increasing community size decreases stability for the systems with fixed connectance ($C$) and a fixed proportion of strong density regulation ($p$). For the maximum ratio of strong density regulation (i.e., for $p=1$) and fixed connectance, systems are always stable for any community size. Fig. (\ref{fig:my_label3}(c),(d)) reveals that the increasing $p$ increases resilience, which means an asymptotic stable system with more strong density regulated species absorbs small perturbation and returns to the equilibrium more quickly. Connectance shows a consistent trend in increasing the resilience of the system for fixed community size (Fig. \ref{fig:my_label3}(c)) as in the case for stability (Fig. \ref{fig:my_label3}(a)). In the case of different community sizes, systems with all species strong density regulated (i.e., for $p=1$) are stable (Fig. \ref{fig:my_label3}(b)), but their absorption rate may differ according to community size (Fig. \ref{fig:my_label3}(d)). Although community size decreases community stability, community size dramatically increases resilience after some threshold proportion for stable systems.
\newline
 
\begin{figure}[H]
    \centering
    \includegraphics[width=12cm]{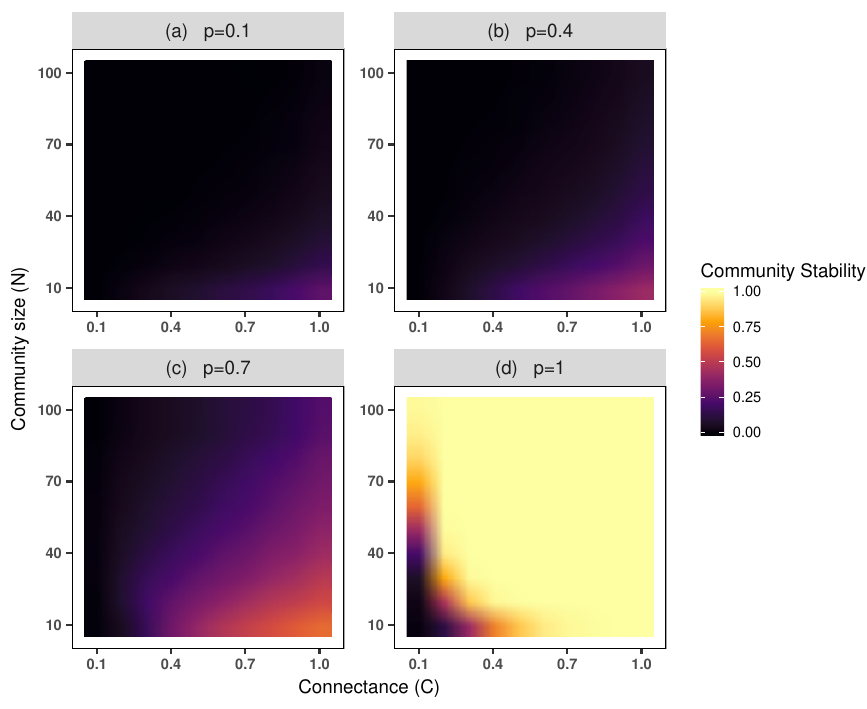}
    \caption{ This panel of figures shows the complexity-stability relationship with increasing the proportion ($p$) of $\theta<1$ species. Stability increases with connectance ($C$). For a low proportion ($p$) of strongly density-regulated species, stability decreases with richness ($N$), which is altered for a high proportion ($p$). Overall stability increases with $p$. }
    \label{fig:my_label4}
\end{figure}

Further Fig. \ref{fig:my_label4}(b) demonstrates the stability complexity relationships for different proportions of $p$. At a low proportion $p$, the stabilizing effect of the connectance and destabilizing effect of community size are investigated. The stabilizing effect of connectance is not too strong for the low proportion of strong density regulation, but the destabilizing of community size is very strong at a low proportion ($p=0.1, 0.4, 0.7$). At the highest proportion $p$ (Fig. \ref{fig:my_label4}(d)), community size acts as a stabilizing factor.\newline

\begin{figure}[H]
    \hspace*{0cm} 
     \centering
     \begin{subfigure}[b]{0.4\textwidth}
         \centering
         \includegraphics[width=\textwidth]{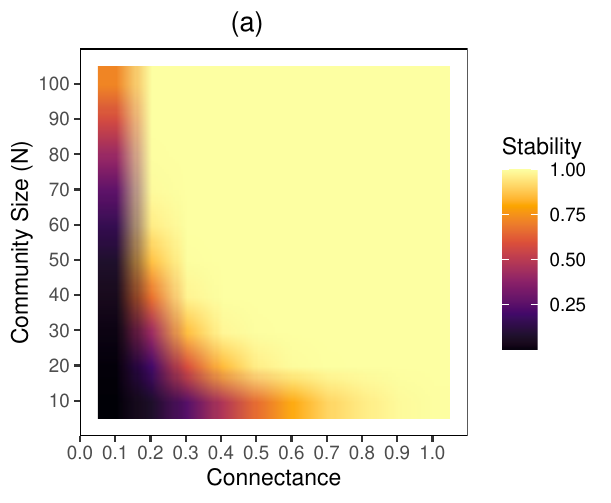}

     \end{subfigure}
     \hspace*{-2cm}
     \hfill
     \begin{subfigure}[b]{0.45\textwidth}
         \centering
         \includegraphics[width=\textwidth]{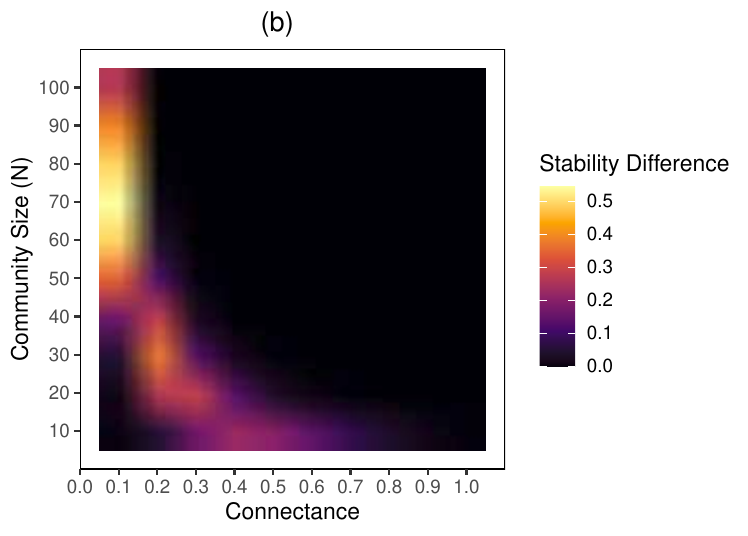}

     \end{subfigure}
    \caption{Panel (a) depicts the complexity-stability relationship with all species exhibiting linear intra-specific growth (i.e., $\theta=1$). Panel (b) shows the difference in stability between species with all concave intra-specific growth and species with all linear intra-specific growth (i.e., $\theta<1$ and $\theta=1$). }
    \label{fig:my_label5}
\end{figure}

We also perform numerical simulation for the case of linear intra-specific density regulation ($\theta=1$), which can be visualized by Fig. \ref{fig:my_label5}(a). For linear density dependence, both community size and connectance stabilize the system. In other words, with increasing community size and connectance, the probability of finding a stable system increases as in the case with $p=1$ (Fig. \ref{fig:my_label4}(d)). However, the stabilization of systems with concave density-dependent ($\theta<1$) is more probable compared to the linear density-dependent systems (Fig. \ref{fig:my_label5}(b)).\newline

\begin{figure}[H]
    \centering
    \includegraphics[width=12cm]{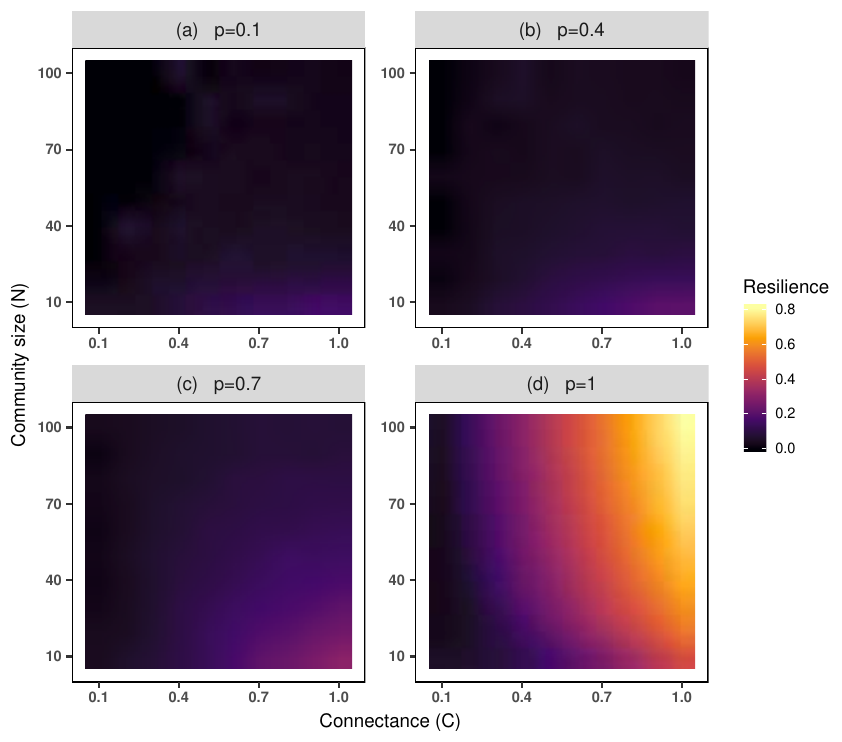}
    \caption{
    This panel of figures shows the complexity-resilience relationship with increasing the proportion $p$ of $\theta<1$ species. Resilience increases with the proportion (p)}. 
    \label{fig:my_label6}
\end{figure}

Similarly, figure \ref{fig:my_label6}(a,c) depicts how the complexity and proportion of strong density regulation affect the community's resilience. At a high proportion ($p=1$), community size enhances the resilience.\newline

\begin{figure}[H]
    \centering
    \includegraphics[width=12cm]{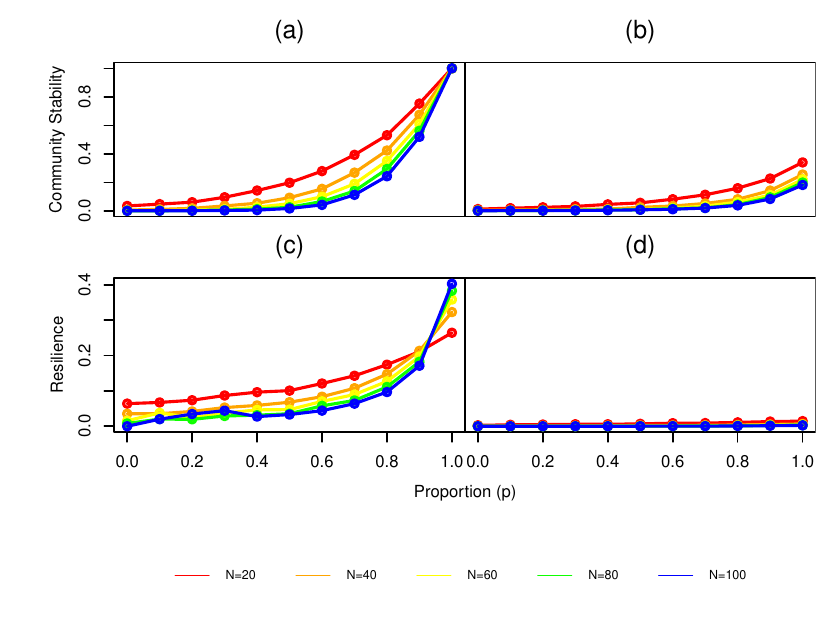}
    \caption{ 
    Panel (a) and (c) represent the community stability and resilience profile for the small world (rewiring probability 0.01) network structure with varying species richness ($N$). The same protocol is followed to generate panels (b) and (d) for the scale-free networks (of power degree exponent 2.1). Community stability and resilience increase with p. Overall stability and resilience are less in the case of a scale-free network compared to the small-world network structure.
 }
    \label{fig:my_label7}
\end{figure}

We also verify the above result for other network types. Fig. \ref{fig:my_label7} demonstrates how strong and weak density regulation affects stability and resilience in scale-free and small-world network structures for various community sizes. For small-world network structures (Fig. \ref{fig:my_label7}(a,c)), stability and resilience are similar to the random network case. But the scale-free network (Fig. \ref{fig:my_label7}(b,d)) exhibits, on average, low community stability and resilience because a scale-free network generates a low connectance network. At the highest proportion ($p=1$), the resilience and community stability decrease for increasing community size, which does not contradict the above observation that community size acts as a stabilizing factor for the systems with $p=1$. In a scale-free network, connectance decreases rapidly by increasing community size, and this lower connectance suppresses the stabilizing effect of community size at the highest proportion($p=1$). \newline

\subsubsection{Case \mbox{II}: No distributional assumption}
\label{No distributional assumption}

In the previous analysis, we have generated equilibrium points from a uniform distribution in such a way that none of the species will go extinct. However, this may not be the case in this scenario, where some species may become extinct depending on initial conditions. It is meaningless to study persistence as a measure of stability in the previous case. But in this scenario, persistence is a natural choice as a measure of stability.  However, in this case, evaluating community stability is not meaningful. We have already excluded the extinct species from the community matrix. So, the eigenvalue spectrum will be on the negative side of the real axis, and hence, the community will always be (local) asymptotically stable.\\

\begin{figure}
    \hspace*{-1.5cm}
    \includegraphics[width=20cm]{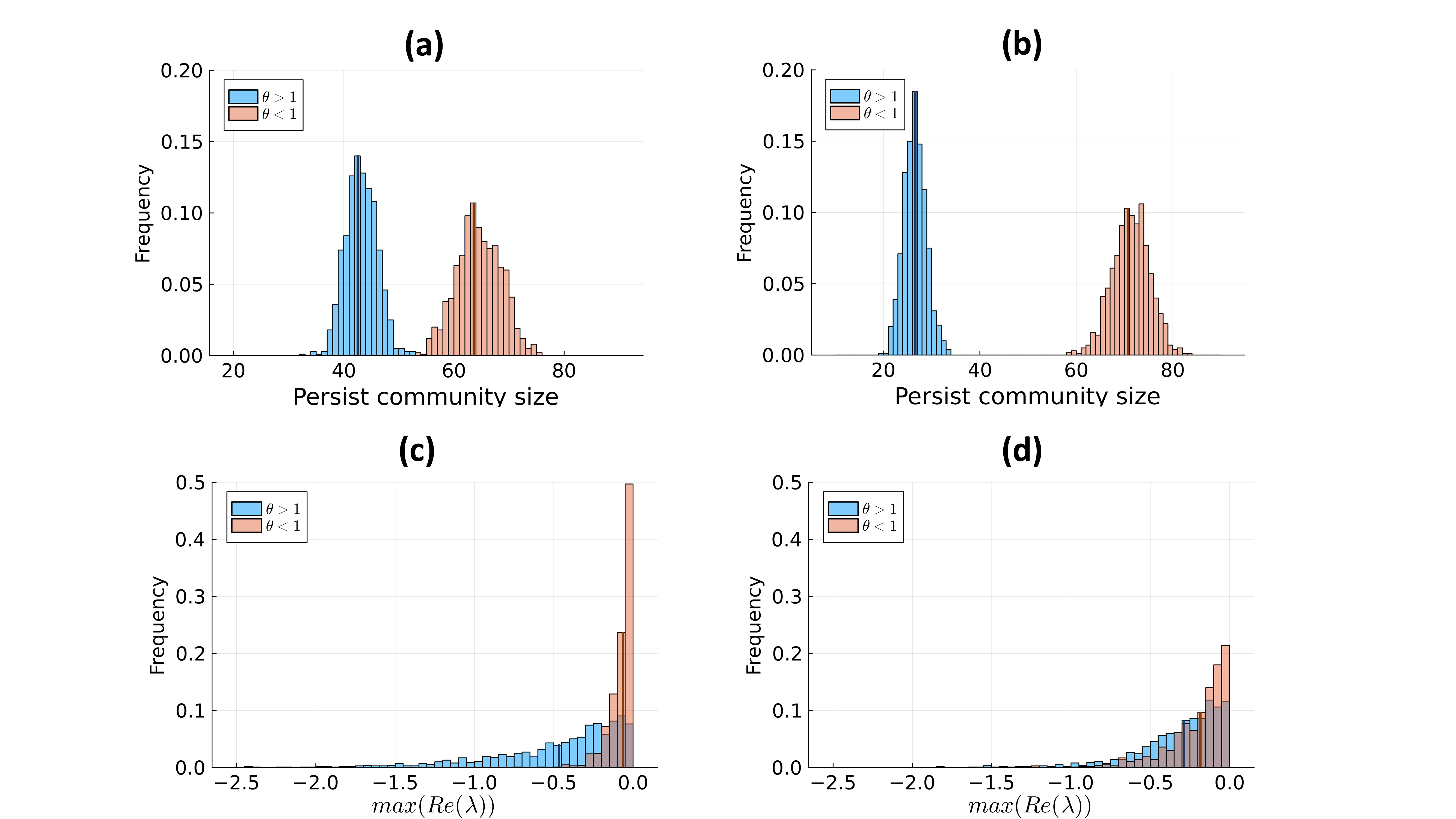}
    \caption{Number of persisting species (a,b) and the distribution of maximum real part of the eigenvalues (c,d) of the persisting community for uncorrelated and negatively correlated $r_{i}$, $\theta_{i}$. where $r_{i}=\gamma/{\theta_{i}}^{\alpha}$ and $\gamma$ is a uniform random variable in $U[0.1,20]$. In (a,c) $\alpha=0$ and in (b,d) $\alpha=1$. Vertical lines denote the mean of the distributions. We set $C=1$ and chose $a_{i,j}$ from $U[-1,0]$. }
    \label{fig:dynamic}
\end{figure}

Figure \ref{fig:dynamic} with four panels illustrates the distribution of the community sizes at equilibrium based on the number of persistent species and the maximum real part of the eigenvalue spectrum of the same equilibrium communities. Fig. \ref{fig:dynamic}(a) and Fig. \ref{fig:dynamic}(c) represent the case where $r$ and $\theta$ are uncorrelated, whereas Fig. \ref{fig:dynamic}(b) and Fig. \ref{fig:dynamic} (d) represent the case when they are correlated. The persistence of the communities with all strong density-regulated species ($p=1$ case) is higher in comparison with the weak density-regulated species ($p=0$ case), irrespective of the correlation structure between $r$ and $\theta$ (Fig. \ref{fig:dynamic}(a,b)). It is important to mention that the disparity between the persistence of the two types of communities ($p=0$ and $p=1$) is more prominent for the case of a negative correlation between $r$ and $\theta$ (histogram of $p=1$ case shifted to the right and another shifted towards left). However, in the case of resilience, we observed an opposite trend compared to our previous analysis. In the case of strong density regulated communities ($p=1$), the maximum real part of the eigenvalues lies near the origin (see Fig. \ref{fig:dynamic}(c)) and hence exhibits less resilience. On the other hand, communities with $p=0$ have that maximum real part far from the origin and hence exhibit more resilience. This disparity between the distribution of the maximum real parts of eigenvalues (i.e., resilience) of the two classes of communities becomes minimal when the $r$ and $\theta$ are negatively correlated (Fig \ref{fig:dynamic}(d)). In this case, we can conclude that the resilience of the two communities is pretty close to each other. \\

\section{Discussion} 
\label{Discussion}

{\it \bf Preface:} It is still an unsettled debate whether the increasing complexity of competitive interactive networks promotes stability in discrete or continuous setups. These competitive networks are based on intra and interspecific competitions. \cite{connell1983evidence} conducted 527 field experiments covering 250 species to determine of prevalence and relative importance of inter-specific competition. In most experiments, the author faced difficulties distinguishing the intraspecific and inter-specific competitions. However, the overall conclusion suggests that marine organisms consistently show higher interspecific competition than terrestrial ones. \newline

A seminal study \citep{may1972will}theorized that the complexity of an ecological community destabilizes the system. Complexity promotes stability under a discrete setup when the selection of competitive partners and strength of interactions are random \citep{fowler2009increasing}. In a continuous setup, a similar phenomenon is observed when we choose the partners randomly, with fixed strength of interactions \citep{rozdilsky2001complexity}. Nevertheless, when both are allowed to vary randomly under the same setup, a reverse phenomenon has been observed \citep{allesina2012stability}.\newline

{\it \bf Density regulation, intra- and inter-specific competition and stability:} \cite{kawatsu2018density} explored the importance of density-dependent interspecific interactions in a mixture network. This network is associated with antagonistic, competitive, and mutualistic relationships. The author classified the interspecific effects as (i) harmful and (ii) beneficial. Density dependence and regulation have a direct impact on species' birth and death rates as components of survivorship in multispecies competitive dynamics. If the density-dependence effect on intraspecific competition is one of the fundamental yardsticks, additionally, with the density-dependence interspecific competition, there must be flexibility in choosing the functional form. This functional form is the natural propensity of the species' growth, which is technically called the intraspecific effect. Improper selection of intraspecific density dependence may lead the entire network to a wrong interpretation of sustainability. These nonlinear forms of interactions can dramatically affect the complexity-stability relationship. The lack of precise understanding of intra- and inter-specific competitions motivated the previous authors to study the theoretical background and mathematical models of the complex multispecies competitive network. \newline

Intraspecific density dependence is important in regulating the long-term species persistence, which has been observed in experimental field studies of insects and fish \citep{amundsen2007intraspecific,arranz2015density}. Density dependence is captured based on the interrelationship between population sizes and food acquisition issues, whereas the link between body size and density dependence is demonstrated in another study (\cite{arranz2015density}). The intraspecific competition must exceed interspecific competition for coexistence (\cite{chesson2013species}). \cite{barabas2016effect} generalizes this well-known criterion. Intraspecific effects have the most stabilizing effect when the species exposed with the largest variance in interspecific interactions. The largest variance is assigned through strong intraspecific coefficients. The authors found that stability requires a large proportion of species with negative self-effect (\cite{barabas2017self}). Intraspecific interference, cannibalism, time scale, the separation between consumers and their resources, spatial heterogeneity, and nonlinear functional responses coupling with their prey are the sources of self-regulation.
So, in a nutshell, density-dependence is a key feature that characterizes the interspecific and intraspecific effects \citep{kawatsu2018density}, and density-dependent variation affects the complexity-stability relationship.  \newline

A complexity-stability relationship emerged for varying species richness and connectance. When the species growth equation involves a nonlinear term, stability may not increase monotonically with species richness \citep{kawatsu2018density}. Previous analysis of \cite{mougi2012diversity} finds that community size destabilizes the competitive community, but the seminal analysis of \cite{fowler2009increasing} finds that species richness stabilizes competitive community. \newline

It might be a fundamental question to ecologists how the stability effect changes with a change of density dependence. More specifically, when the theta logistic pattern governs the density regulation, what will be the consequence in the mode of stability for the above complex competitive network? The existing studies on competitive multispecies networks assumed the linear density-dependence to explain intra-specific competition, which leads to the common logistic law. In this context, \cite{hanski1990density} concluded that the density dependence, regulation, and variability enhance the system complexity.   \newline

The present study introduces nonlinear growth in community-level dynamics by considering nonlinear intraspecific density regulation in species growth profiles. We initiate our analysis by deriving the theoretical conditions for stability, which involve the ratio of species exhibiting weak and strong self-regulation during initial growth phases, as well as the ecological network's complexity across a competitive community ensemble. We find that stability increases as the relative proportion of species that strongly self-regulate at low population density increases compared to at high population density (see  Fig. \ref{fig:my_label 1}). The proportion of such species in the systems affects stability through the eigenvalue pushback effect and Jacobian averaging effect \citep{gravel2016stability}. We perform numerical simulations to justify our results by assuming that every species coexists at the equilibrium. According to our results, competitive webs with strong density-regulated species are more stable because the coexistence probability is relatively higher than the systems containing fewer species(see Fig. \ref{fig:my_label3}). Region of stability also increases in the case of $\theta<1$ compared to $\theta=1$ (see Fig. \ref{fig:my_label5}). Moreover, the findings indicate that communities with a higher proportion of strongly density-regulated species exhibit greater resilience (Fig. \ref{fig:my_label3}(c, d)). Conversely, in scenarios where species' equilibrium abundances are not presumed, and extinctions may occur over time, persistence follows a similar pattern. Specifically, persistence, which is considered a trivial measure of stability \citep{landi2018complexity}, increases with the presence of strongly density-regulated species (Fig. \ref{fig:dynamic}). In instances of correlated intrinsic growth and density-regulation parameters, communities featuring strong density-regulated species demonstrate comparable resilience to those with weak density-regulation despite significant variations in persistent community sizes. Given the highly reproductive nature of such strong density-regulated species \citep{clark2010theta, saether2002demographic, saether2008forms, fowler1988population}, they can swiftly recover from moderate abundance reductions due to their high reproductive ability, returning to equilibrium abundances faster than other species \citep{krebs2009behavioural}.
\newline

In this study, we have considered three types of network topology to validate the robustness of our result. Here, we intuitively describe the role of these three network types from the aspect of a competitive community. First, we consider a random network structure where any species has an equal probability of competing with others in the community. So, the competitive links between species are generated randomly with some specified probability (connectance) \citep{chen2001transient}. Secondly, we consider a scenario where a large number of species may face only a few competitors while a few species may encounter many competitors. These can be identified as specialist and generalist species, respectively. Generalist species typically possess broader niche widths, allowing them to share their niche and form competitive links with more species than specialist species with narrower niche widths \citep{may1972niche}. To accommodate this phenomenon, we adopt a scale-free network structure where most nodes have few links, and only a few nodes have many connections. Lastly, we consider small-world networks, known for their high clustering coefficients \citep{newman2018networks}. The clustering coefficient of a node indicates how closely the neighbors of a node are connected. Now, if two species have a common competitor, then it is most likely that they will also compete with each other. Our results are consistent with all the above network architectures. \newline

{\it \bf Interconnection of r, K-selection and density regulation:} The quintessence of the concept of r, K-selection is that organisms toil to maximize their fitness for survival in either uncrowded (r-selection) or crowded (K-selection) environments. Fitness is conventionally defined as the proportion of genes left in the population gene pool (\cite{pianka2011evolutionary}, page 10). 
Fitness is a poorly developed concept with fuzziness in definition \citep{mcgraw1996estimation}. More importantly, components of fitness are inadequate surrogates for individual fitness in the age structure population of birds. In fact, several recent ecological studies suggest fitness depends more heavily on annual survivorship than annual fecundity for most taxa \citep{crone2001survivorship}. So, there is a scope for ecologists to define fitness from various perspectives. \newline

Depending on different definitions of the fitness function, several interpretations of r, K-selection theory can come out for the ecologists. As per the classical definition, r, K-selection theory is generally explained by the survivorship curve. This curve plots the survivorship function against time (age). Here, the survivorship function is defined as the logarithm of the ratio of the populations that survive from one age to the next. In other words, survivorship is the number of individuals in a population that can be expected to survive to any specific age, hence the population's fitness. \newline

Note that the per-capita growth rate \citep{fisher1930genetical} is the Malthusian parameter of Darwinian fitness. This fitness is either a linear (logistic) or nonlinear power (theta-logistic) function of the per-capita growth rate. As suggested by \cite{clark2010theta}, concave r-N curves are typical of so-called r-selected organisms where density dependence acts strongly at lower densities. In contrast, convex curves arise from K-selected species, where density dependence acts to reduce growth only at higher densities. \newline

The mean individual fitness is measured as per-capita growth rate while defining the Allee effect in biology. Populations exhibiting a weak Allee effect will possess reduced individual fitness at a lower population size. However, the population will always exhibit a positive per-capita growth rate, which is different from a strong Allee effect, where the pgr becomes negative below a critical population size. So, pgr is an excellent surrogate for defining fitness. So, the life history pattern can be derived with the density-pgr relationship as described in \cite{williams2013accounting} (page-6). r-selected species are expected to depict a profile of strong density-dependent reduction in small populations because of more significant consumption of resources by increasing reproduction output. K-selected species shows the opposite response by producing a pattern of weak density-dependent reduction at smaller population \citep{fowler1981density,saether2002pattern,nicoll2003declining,mcghee2007estimation,coulson2008estimating}. \newline
 
 The growth response of the theta-logistic model governed by the density-pgr curve ideally characterizes a population unable to recover quickly from extrinsic perturbations when $\theta<1$ (strong density regulation), i.e., concavity presence in the profile. On the other hand, a convex growth response ($\theta>1$) implies that density feedback occurs mainly above some relatively large threshold population abundance. So, the theta-logistic model can be focused around r vs. K-selected species based on $\theta$ $<$ and $>$ 1, respectively. The classification for the multispecies competitive model, where the theta-logistic model regulates the intraspecific competition, is a bit obscure. Further study is necessary to establish this theory as a pedagogical metaphor. If we assume this as a pedagogical metaphor, the analysis based on our proposed model suggests that the stability of the competitive interactive network is enhanced if the r-selected species are dominated in society. These findings are not unrealistic. The findings support the \cite{sibly2005regulation} findings of concave upward density-pgr relationship dominance. \newline
 
\begin{figure}[H]
    \centering
    \includegraphics[width=14cm]{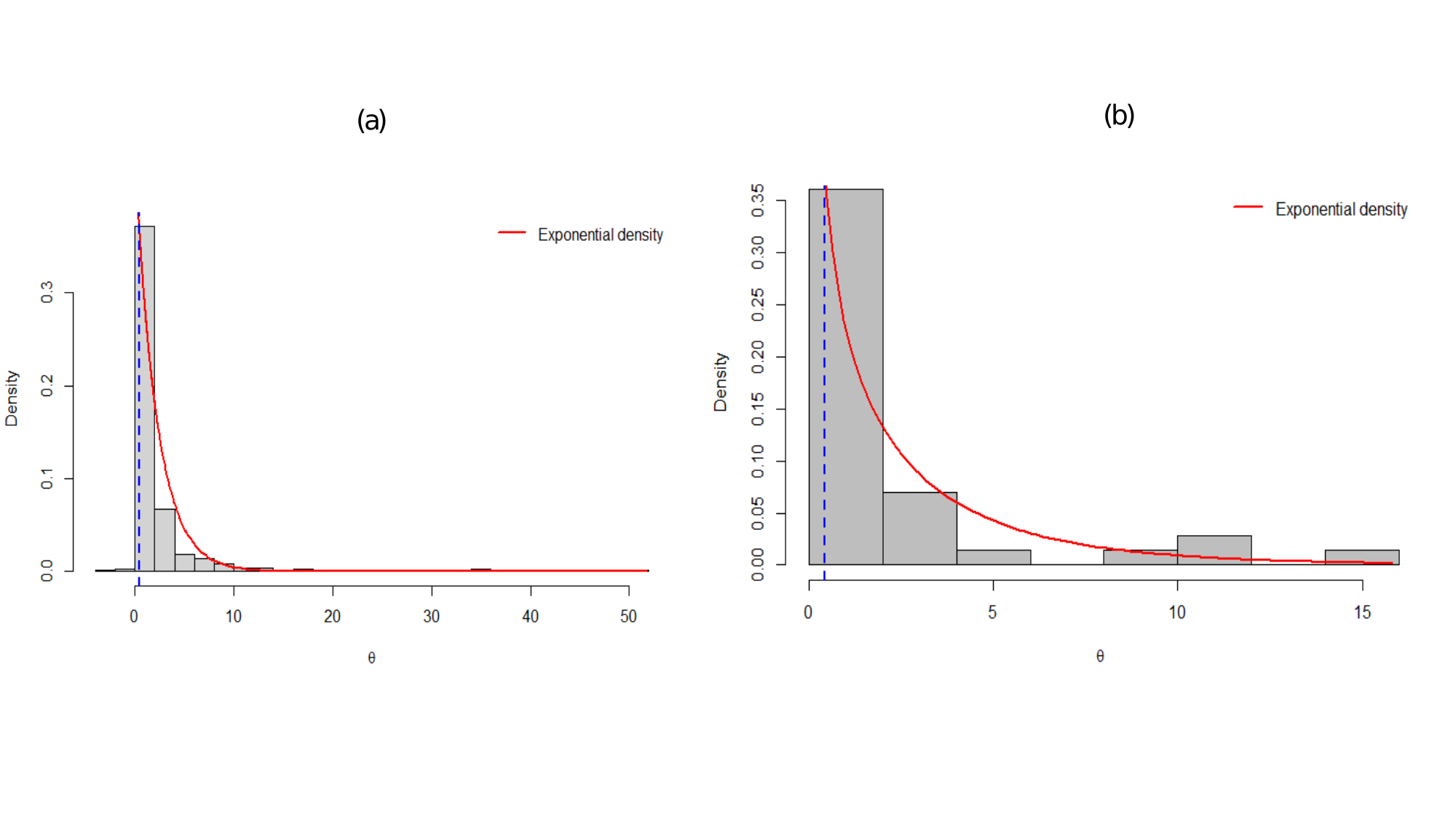}
    \caption{The histogram of two probability densities of the density regulation parameter $\theta$ of the theta-logistic equation from $5000$ time series of the GPDD database. Panel (a) and (b) represent the histogram of $\theta$, without and with the positivity restriction. The red curve represents the corresponding exponential density curve, and the blue line represents the corresponding mean, which is $0.43$ and $0.45$, respectively. }
    \label{fig:my_label8}
\end{figure}

{\it \bf Limitation and future direction:} The literature already described in different parts of the manuscript on density-dependent competitive network randomization is inserted either in the selection of competitive partners or in the strength of interaction. 
Instantaneous growth rate (IGR) in a density-regulated setup is the most vital parameter, which has been affected by demographic stochasticity (\cite{loeschcke1991species}). There are also many instances where IGR is affected by environmental and demographic stochasticity (\cite{sinervo2010erosion}). This randomized version of IGR will give birth to the new density-pgr relationship. This new connection must impact the community stability of the complex network. The vital criticism from \cite{sibly2005regulation}'s study on the topic of selection of impresses parameter estimation method. In many cases, \cite{sibly2005regulation} estimated "$\theta$" to be negative, and that leads to ecologically implausible estimates of "r." \cite{clark2010theta} disagree with the final conclusion of \cite{sibly2005regulation}'s analysis. Mainly for neglecting the measurement errors. He raised questions on the reliability of the theta-logistic model. We believe the issue should be unbiasedly judged. The problem is not the model but the estimation method. Profile likelihood is the best method of parameter estimation in this case. We put a significant effort into showing the performance of the profile likelihood method, in this case, using the 5000-time series from GPDD \citep{nerc2010global}. We showed that the profile hardly generates negative estimates of r under a stochastic setup. We adopted the method of \cite{paul2022estimation} for this test-bed analysis and generated two histograms of estimated $\theta$. The probability density is fitted with the exponential density with the sample mean as the estimate for the parameter mean. The estimated mean of the two probability densities with and without positive restriction on $\theta$, 0.45 and 0.43 (Figure \ref{fig:my_label8}), respectively. However, it needs more rigorous mathematical meanderings in the form of Ito calculus. This new theoretical aspect might be an exciting research arena for future endeavors.

\section{Conclusions}
\label{Conclusions}

Understanding the functional form of density dependence associated with life-history strategies and its interaction with other species is crucial to shaping our ability to predict ecosystem stability more precisely. The work presented here considers the contribution of nonlinear density regulation to the stability of the competitive community. Interestingly, our work reveals that complex, competitive communities are stable if the relative ratio of species exhibiting strong intraspecific density-dependent regulation at low population density (concave self-regulation) outweighs those at high population density (convex self-regulation). Alternatively, increasing r-selected species in a competitive ecological community can enhance the system's stability. This study can provide clues for the sustainability of complex ecological networks and the maintenance of the population biodiversity. We call for future empirical studies on density-dependent growth profiles to validate our results' truthfulness in nature.

\section*{Acknowledgement}
Amit Samadder acknowledges the Senior Research Fellowship (file no: 09/093(0189)/2019-EMR-I) from the Council of Scientific and Industrial Research (CSIR), India. Arnab Chattopadhyay acknowledges a Senior Research Fellowship (file no: 09/093(0190)/2019-EMR-I) from the Council of Scientific and Industrial Research (CSIR), India. Anurag Sau would like to acknowledge the Senior Research Fellowship from UGC, India, which funded his work. We are thankful to Mr. Ayan Paul for providing some data. Additionally, we extend our gratitude to the editor and the reviewer for their insightful comments, which greatly contributed to improving the manuscript's quality.

\section{Author contributions} {\bf Amit Samadder}:  Conceptualization, model formulation,  method, numerical simulation, manuscript writing, and editing; {\bf Arnab Chattopadhyay}: Model formulation, numerical simulation, result validation, manuscript writing, and editing; {\bf Anurag Sau}: Manuscript writing and editing; {\bf Sabyasachi Bhattacharya}: Supervision, manuscript writing, and editing. 

\section*{Conflict of interest} The authors declare no competing interests.

\newpage
\begin{appendices}

\section{Derivation of the condition for stability}\label{Appendix:A}

Let, $M$ is the community matrix with connectance $C$, i.e, $M_{ij}<0$ with probability $C$ and $M_{ij}=0$ with probability $1-C$. Further, we assume that there are two types of self-regulating terms $d_{1}$ and $d_{2}$ in $M$ with probability $P$ and $1-P$, i.e., each diagonal entry of $M$ takes the value $d_{1}$ with probability $P$, $d_{2}$ with probability $1-P$.
 \newline
 
 Let, the eigenvalues of $M$ be $\lambda_{i} (i=1, 2, ..., N)$. The expected spectral distribution of an ensemble of matrices can be written as $w(z)=E(\delta(z-\lambda_{i}))$ where $E(.)$ denotes expectation, and $\delta(.)$ denotes the Dirac delta function.
 \newline
 
 Local stability of the equilibrium point $X^{*}$ depends on the spectral density abscissa $\eta(M)$ of $ M.$ Here $\eta(M)=sup\{Re(z):z\in support(w)\}$. The condition for the local stability of the system \ref{final_equation} in the main text is $\eta(M)$ should be less than zero.

 \subsection*{Spectral Density abscissa of competitive community matrix}
 \label{Spectral Density abscissa of competitive community matrix}
 
 \begin{theorem}
 Let $M$ be a competitive community matrix with $N$ species and connectance C, whose off-diagonal entries are sampled from the distribution $-|\mathbf{X}|$ with probability $C$ and 0 with probability $1-C$, where $E(\mathbf{X})=0$, $Var(\mathbf{X})=\sigma^{2}$i.e, $E(M_{ij})_{i\neq j}=-CE(|\mathbf{X}|)$, $Var(M_{ij})_{i\neq j}=C\sigma^{2}-C^{2}E^{2}(|\mathbf{X}|)$. Let, the diagonal entries follows the distribution $p(h)=P\delta(h-d_{1})+(1-P)\delta(h-d_{2})$, $\Delta=d_{2}-d_{1}>0$.
 \newline
 Then the spectral density abscissa  $\eta(M)$ is:
 {\small
 \begin{equation}
 \eta(M)=x{\sqrt{N(C\sigma^{2}-C^{2}{E^{2}}(|\mathbf{X}|))}} +d_{2}+\frac{{\sqrt{N}}C(1-C){E^{2}}(|\mathbf{X}|)}{{\sqrt{(C\sigma^{2}-C^{2}{E^{2}}(|\mathbf{X}|))}}}\left({\frac{P}{x+\frac{\Delta}{\sqrt{N(C\sigma^{2}-C^{2}{E^{2}}(|\mathbf{X}|))}}}}+\frac{1-P}{x}\right)+CE(|\mathbf{X}|)
 \end{equation}}
 for every $0\leq P\leq1$, $x$ satisfying
 \begin{equation}
  1=\frac{P}{(x+\frac{\Delta}{\sqrt{N(C\sigma^{2}-C^{2}{E^{2}}(|\mathbf{X}|)}})^{2}}+\frac{1-P}{x^{2}} 
\end{equation}
\end{theorem}

\textbf{Proof:} Let, $M^{'}$ be a $N{\times}N$ standard elliptic matrix with diagonal entries $M_{ii}^{'}=0$ , $E(M_{ij}^{'})_{i \neq j}=0, E({M_{ij}^{'}}^{2})_{i \neq j}=\frac{1}{N}$ and $E({M_{ij}}^{'}{M_{ji}^{'}})_{i \neq j}=\frac{\rho}{N}$.
\newline 

$D^{'}$ is a $N{\times}N$diagonal matrix with diagonal entries from
\begin{equation*}
 p(h)=P\delta(h-d_{1})+(1-P)\delta(h-d_{2}),   
\end{equation*}
where $\delta(.)$ is the Dirac delta function. In other words, any diagonal entry will be set equal to $d_{1}$ with
probability $P$ and to $d_{2}$ with probability $1-P$. Since $D^{'}$ is diagonal, its diagonal entries are also
equal to its eigenvalues. We assume that $\Delta=d_{2}-d_{1}>0$ and $M^{'}+D^{'}=A^{'}$.
\newline

According to \cite{barabas2017self} the spectral abscissa of $A^{'}$ is
\begin{equation*}
    \eta(A^{'})=x+d_{2}+\rho\left(\frac{P}{x+\Delta}+\frac{1-P}{x}\right),
\end{equation*}
where
\begin{equation*}
     1=\frac{P}{(x+\Delta)^{2}}+\frac{1-P}{x^{2}}.
\end{equation*}

We extend it for the case of the non-elliptic matrix  $Q$ ($=M-D^{'}$), the non-diagonal part of $M$. 
\newline

We can write $M$ as,
\begin{equation*}
    M=(Q-CE(\mathbf{|X|})I+CE(\mathbf{|X|})F)+CE(\mathbf{|X|})I-CE(\mathbf{|X|})F+D^{'},
\end{equation*}
where, $F$ is a square matrix of order $N$ with all entries equal to $1$. \\
By substituting  $S=Q-CE(\mathbf{X})I+CE(\mathbf{|X|})F$ we get,
\begin{equation*}
    M=(S+D^{'})+CE(\mathbf{|X|})I-CE(\mathbf{|X|})F,
\end{equation*}
 Where $E(S_{ij})_{i \neq j}=0$, $E(S_{ij}^{2})_{i \neq j}=C\sigma^{2}-C^{2}{E^{2}}(|\mathbf{X}|)$.
\newline

We define
\begin{equation*}
    S^{'}=\frac{S}{\sqrt{N(C\sigma^{2}-C^{2}{E^{2}}({|\mathbf{X}|}))}}
\end{equation*}
Then, $E(S_{ij}^{'})_{i\neq j}=0$, $E({S_{ij}^{'}}^{2})_{i \neq j}=\frac{1}{N}$ and $E(S_{ij}^{'}S_{ji}^{'})_{i \neq j}=\frac{\rho}{N}$, where $\rho=\frac{C(1-C){E^{2}}(|\mathbf{X}|)}{C\sigma^{2}-C^{2}{E^{2}{(|\mathbf{X}|)}}}$. So, $S^{'}$ is the standard elliptic matrix and hence $S^{'}+\frac{D^{'}}{\sqrt{N(C\sigma^{2}-C^{2}{E^{2}}({|\mathbf{X}|}))}}$ has the spectral abscissa: 

\begin{multline*}
    \eta(S^{'}+\frac{D^{'}}{\sqrt{N(C\sigma^{2}-C^{2}{E^{2}}({|\mathbf{X}|}))}}) =x+\frac{d_{2}}{\sqrt{N(C\sigma^{2}-C^{2}{E^{2}}(|\mathbf{X}|))}}+\\
   \frac{C(1-C){E^{2}}(|\mathbf{X}|)}{C\sigma^{2}-{C^{2}}{E^{2}}(|\mathbf{X}|)}\left(\frac{P}{x+\frac{\Delta}{\sqrt{N(C\sigma^{2}-C^{2}E^{2}(|\mathbf{X}|))}}}+\frac{1-P}{x}\right),
\end{multline*}

\begin{equation*}
  \frac{P}{(x+\frac{\Delta}{\sqrt{N(C\sigma^{2}-C^{2}{E^{2}}(|\mathbf{X}|))}})^{2}}+\frac{1-P}{x^{2}}=1.     
\end{equation*} 
\\

 In terms of $S^{'}$ the expression of $M$ can be written as,
\newline
$M=(S+D^{'})+CE(\mathbf{|X|})I-CE(\mathbf{|X|})F\\={\sqrt{N(C\sigma^{2}-C^{2}{E^{2}}({|\mathbf{X}|}))}}\left(S^{'}+\frac{D^{'}}{\sqrt{N(C\sigma^{2}-C^{2}{E^{2}}({|\mathbf{X}|}))}}\right)+CE(|\mathbf{X}|)I-CE(|\mathbf{X}|)F$.
\newline

Following \cite{gibbs2018effect}, the bulk of the eigenvalues of $(S+D^{'})+CE(\mathbf{|X|})I-CE(\mathbf{|X|})F$ is equal to the bulk of the eigenvalue $(S+D^{'})+CE(\mathbf{|X|})I$ and an additional outlier which is approximately equal to sum of the two terms, mean of the distribution $p(h)$ and -$CE(\mathbf{|X|)}(N-1)$, i.e.,  $Pd_{1}+(1-P)d_{2}-CE(\mathbf{|X|})(N-1)$.
\newline

We get the spectral abscissa  of $M$ as,

$\eta(M)=max(\sqrt{N(C\sigma^{2}-C^{2}E^{2}(|\mathbf{X}|))}\eta(S^{'}+\frac{D^{'}}{\sqrt{N(C\sigma^{2}-C^{2}{E^{2}}({|\mathbf{X}|}))}})+CE(|\mathbf{X}|)$, $Pd_{1}+(1-P)d_{2}-CE(\mathbf{|X|})(N-1)$).
\newline

By simplifying, we get,

{\small
\begin{equation*}
\begin{split}
    \eta(M)& =max(x{\sqrt{N(C\sigma^{2}-C^{2}{E^{2}}(|\mathbf{X}|))}} +d_{2}
+\frac{{\sqrt{N}}C(1-C){E^{2}}(|\mathbf{X}|)}{{\sqrt{C\sigma^{2}-C^{2}{E^{2}}(|\mathbf{X}|)}}}\left({\frac{P}{x+\frac{\Delta}{\sqrt{N(C\sigma^{2}-C^{2}{E^{2}}(|\mathbf{X}|))}}}}+\frac{1-P}{x}\right)\\
&+CE(|\mathbf{X}|), Pd_{1}+(1-P)d_{2}-CE(\mathbf{|X|})(N-1))
\end{split}
\end{equation*}
}
for every $0\leq P\leq1$, $x$ satisfying
\begin{equation*}
   \frac{P}{(x+\frac{\Delta}{\sqrt{N(C\sigma^{2}-C^{2}{E^{2}}(|\mathbf{X}|))}})^{2}}+\frac{1-P}{x^{2}}=1.
\end{equation*}
 
 Now we show that when $d_{1},d_{2}$ are negative, the second term within the maximum function is always smaller than the first term.
 \newline
 
 When $d_{1} \xrightarrow{}{d_{2}}$, we get $x=1$ \citep{barabas2017self} ,i.e,\\
 \newline
 $\eta(M)=max({\sqrt{N(C\sigma^{2}-C^{2}{E^{2}}(|\mathbf{X}|))}}+d_{2}+\frac{{\sqrt{N}}C(1-C){E^{2}}(|\mathbf{X}|)}{{\sqrt{(C\sigma^{2}-C^{2}{E^{2}}(|\mathbf{X}|)}}}+CE(|\mathbf{X}|),d_{2}-CE(\mathbf{|X|})(N-1))={\sqrt{N(C\sigma^{2}-C^{2}{E^{2}}(|\mathbf{X}|))}}+d_{2}+\frac{{\sqrt{N}}C(1-C){E^{2}}(|\mathbf{X}|)}{{\sqrt{(C\sigma^{2}-C^{2}{E^{2}}(|\mathbf{X}|)}}}+CE(|\mathbf{X}|)$.
 \newline
 When $0<P<1$,
 \begin{equation*}
    \frac{P}{(x+\frac{\Delta}{\sqrt{N(C\sigma^{2}-C^{2}{E^{2}}(|\mathbf{X}|))}})^{2}}+\frac{1-P}{x^{2}}-1=0 
 \end{equation*}
 \begin{equation*}
     \implies x^{2}P+(x+\frac{\Delta}{\sqrt{N(C\sigma^{2}-C^{2}{E^{2}}(|\mathbf{X}|))}})^{2}(1-P)-x^{2}(x+\frac{\Delta}{\sqrt{N(C\sigma^{2}-C^{2}{E^{2}}(|\mathbf{X}|))}}))^{2}=0.
 \end{equation*}
\\
 
 Assuming,
 \begin{equation*}
    f(x)=x^{2}P+(x+\frac{\Delta}{\sqrt{N(C\sigma^{2}-C^{2}{E^{2}}(|\mathbf{X}|))}})^{2}(1-P)-x^{2}(x+\frac{\Delta}{\sqrt{N(C\sigma^{2}-C^{2}{E^{2}}(|\mathbf{X}|))}}))^{2}
 \end{equation*}
 we have 
 $f(1)=P\left(1-(1+\frac{\Delta}{\sqrt{N(C\sigma^{2}-C^{2}{E^{2}}(|\mathbf{X}|))}})^{2} \right)<0$, and $f(0)=(1-P)(\frac{\Delta}{\sqrt{N(C\sigma^{2}-C^{2}{E^{2}}(|\mathbf{X}|))}})^{2}>0$.
 \newline
 
 According to the intermediate value property of a continuous function, $f(x)$ has at least one root in $(0,1)$. Thus, for $0<P<1$, we get $0<x<1$, so the first term within the maximum function is positive except $d_{2}$ and    
 $Pd_{1}+(1-P)d_{2}-CE(\mathbf{|X|})(N-1)=-P\Delta+d_{2}-CE(\mathbf{|X|})(N-1)<d_{2}$. Hence, the second term within the maximum function is smaller than the first term. One can write,
\newline

\begin{equation*}
\begin{split}
  \eta(M)&=x{\sqrt{N(C\sigma^{2}-C^{2}{E^{2}}(|\mathbf{X}|))}} +d_{2}
+\frac{{\sqrt{N}}C(1-C){E^{2}}(|\mathbf{X}|)}{{\sqrt{(C\sigma^{2}-C^{2}{E^{2}}(|\mathbf{X}|)}}}\left({\frac{P}{x+\frac{\Delta}{\sqrt{N(C\sigma^{2}-C^{2}{E^{2}}(|\mathbf{X}|))}}}}+\frac{1-P}{x}\right)\\
&+CE(|\mathbf{X}|)  
\end{split}
\end{equation*}

for every $0\leq P\leq1$, $x$ satisfying

\begin{equation*}
   1=\frac{P}{(x+\frac{\Delta}{\sqrt{N(C\sigma^{2}-C^{2}{E^{2}}(|\mathbf{X}|))}})^{2}}+\frac{1-P}{x^{2}}.
\end{equation*}

 \end{appendices}

 \bibliographystyle{elsarticle-harv} 
\bibliography{bibliography.bib}

\end{document}